\documentclass[twocolumn,showpacs,preprintnumbers,amsmath,amssymb]{revtex4}
\usepackage{graphicx}% Include figure files
\usepackage{dcolumn}% Align table columns on decimal point
\begin{document}

%\preprint{APS/123-QED}

\title{Verifying continuous variable entanglement of intense light pulses}% Force line breaks with \\

\author{Oliver Gl\"ockl}
\email{gloeckl@kerr.physik.uni-erlangen.de}
\author{Ulrik L. Andersen}
\author{Gerd Leuchs}
% \altaffiliation[Also at ]{Physics Department, XYZ University.}%Lines break
%automatically or can be forced with \\
%\author{Second Author}%
% \email{Second.Author@institution.edu}
\affiliation{%
Institut f\"ur Optik, Information und Photonik, Max--Planck Forschungsgruppe,
Universit\"at Erlangen--N\"urnberg,\\
G\"unther--Scharowsky--Stra{\ss}e 1 / Bau 24, 91058 Erlangen, Germany}%

%\author{Charlie Author}
% \homepage{http://www.Second.institution.edu/~Charlie.Author}
%\affiliation{
%Second institution and/or address\\
%This line break forced% with \\
%}%

\date{\today}% It is always \today, today,
             %  but any date may be explicitly specified

\begin{abstract}
Three different methods have been discussed to verify continuous variable
entanglement of intense light beams. We demonstrate all three methods using
the same set--up to facilitate the comparison. The non--linearity used to
generate entanglement is the Kerr--effect in optical fibres. Due to the
brightness of the entangled pulses, standard homodyne detection is not an
appropriate tool for the verification. However, we show that by using large
asymmetric interferometers on each beam individually, two non-commuting
variables can be accessed and the presence of entanglement verified via joint
measurements on the two beams. Alternatively, we witness entanglement by
combining the two beams on a beam splitter that yields certain linear
combinations of quadrature amplitudes which suffice to prove the presence of
entanglement.
\end{abstract}

% PACS, the Physics and Astronomy
                             % Classification Scheme.
%\keywords{Suggested keywords}%Use showkeys class option if keyword
                              %display desired
\maketitle

\section{Introduction}
About 70 years after Einstein described it as "spooky interactions at a
distance"~\cite{EIN35}, the phenomenon of entanglement is one of the
most intriguing and useful properties of quantum mechanics. In the very early
days of entanglement its philosophical mindboogling implications dominated the
discussion; later it served as an experimental testbed for the foundations of
quantum mechanics and recently the usage of entanglement has facilitated a revolution in
information science holding the promise to efficiently solve hard computational
problems and guranteeing secure communication~\cite{NIE00}. More
specifically, entanglement is the key physical resource enabling the successful
execution of many quantum information protocols, prominent examples being
quantum teleportation~\cite{BEN93} and quantum dense
coding~\cite{BEN92}. These protocols were initially proposed for discrete
variables, but recently protocols utilizing continuous quantum variables, such
as the quadrature amplitudes of the electromagnetic field, have emerged as an
interesting alternative~\cite{BRA03}. Therefore over the last few
years there has been a desire to develop efficient and compact sources of
entanglement where quadrature variables of the light state are quantum
mechanically correlated which is normally coined quadrature entanglement.

To date many convincing experimental realisations of quadrature entanglement
have been conducted based on different nonlinear effects in different media such
as the second order nonlinearity in optical parametric oscillators either
utilizing a single nondegenerate OPO (in
polarization~\cite{OU92,ZHA00b,LAU04},
frequency~\cite{SCHO02,VIL05} or direction~\cite{WEN05}) or two
degenerate OPOs~\cite{FUR98,BOW03b,AOK03}, and the third
order nonlinearity in atoms~\cite{JOS04} or
fibers~\cite{SIL01}. After the production of entanglement it is
crucial to have some experimental characterisation methods at hand by which one
can determine whether the system contains truely quantum mechanical correlations.

In all the above mentioned experiments, Gaussian states were assumed to be
produced. In this case complete information about the system, including the
amount of entanglement, is contained in the covariance matrix. However, instead
of measuring all entries of the covariance matrix (which is quite demanding,
experimentally) the variance of certain linear combinations of the quadrature
variables suffice to witness the presence of
entanglement~\cite{REI88,DUA00,SIM00}. Such a criterion for
entanglement was used in all the experiments listed above. In most of the
experiments homodyne detection on each subsystem (relying on two strong local
oscillators) followed by classical communication between them were used to
generate the appropriate linear combinations and subsequently witness the
presence of entanglement. Such an approach was indeed valid in these experiments
since the entangled beams under investigations either did not have an optical
carrier or one that was negligible small compared to that of that of the local
oscillator. However, by exploiting the Kerr effect in fibers to generate
squeezing and subsequently entanglement, pulses containing on the order of
$10^8$ photons or more per pulse are needed to compensate for the almost
vanishing small Kerr nonlinearity in standard glass fibers. In turn this
practically excludes the usage of homodyne detection techniques since this will
require local oscillator pulses containing more than $10^{10}$ photons and hence
detectors capable of handling high power pulses and simultaneously measuring at
the quantum noise limit, a technology which is not, at present, technically
feasible.

In this paper we present a careful verification of the presence of entanglement
between intense pulses by constructing the proper linear combination of
noncommuting quadrature variables by means of three different measurement
strategies. In the  first method we prove entanglement by measuring locally the
amplitude and phase quadratures of the two beam and compare the outcomes using
classical communication. Here the conjugate quadratures of the respective beams
are measured using an interferometric method~\cite{GLO04}.
As an alternative, we verify entanglement between two beams by interfering them
on a beam splitter and then by controlling their relative phase shift, we are
able to measure the correlations of the conjugate quadratures either separately
by comparing the two outputs from the beam splitter or jointly by monitoring a
single output. The latter method was applied in ref.~\cite{SIL01} to
verify the presence of entanglement. Here we elaborate on the work of
ref.~\cite{SIL01} by conducting a more careful analysis of the
different methods by which proper linear combinations of quadrature operators
can be constructed to witness the presence of entanglement.

The paper is structured as follows. In section \ref{sec-production}, we
review how continuous variable entanglement is generated by a linear combination
of two squeezed beams on a beam splitter and how the presence of entanglement
can be detected using a criterion based on quantities that can be
accessed easily in experiments. In section
\ref{sec-experiment}, a description of the experimental setup is outlined.
Section \ref{sec-techniques}, which constitutes the main part of this paper, is
devoted to a careful description of the three different experimental approaches
used to extract the linear combination of quadrature variables that proves the
presence of entanglement. In section \ref{sec-conclusions} we conclude the work.

\section{Entanglement production and verification}\label{sec-production}
\begin{figure}
\includegraphics[scale=.5]{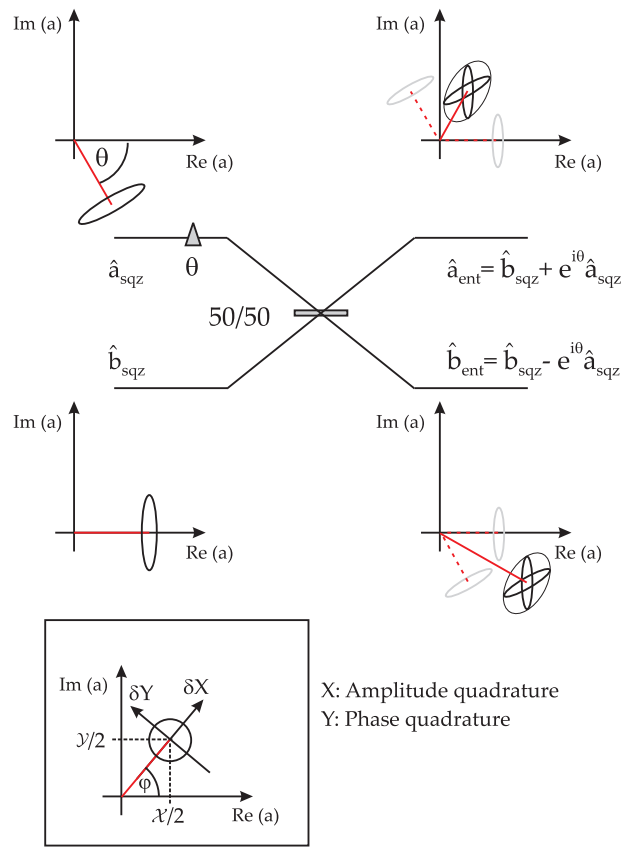}
\caption{\label{fig-ent} Schematic representation of entanglement generation in
a phase space diagram. Two bright amplitude squeezed beams interfere with
relative phase $\theta$ at a 50/50 beamsplitter. Note that the plotted situation
corresponds to non ideal entanglement generation where $\theta\neq\pi/2$, the
resulting uncertainty areas are elliptical and the uncertainty in amplitude and
phase is different for the two modes. Throughout this paper, the amplitude
quadrature $\hat X$ is defined to point along the direction of the classical
excitation, the phase quadrature $\hat Y$ is perpendicular (see inset).}
\end{figure}

For the theoretical description of
squeezing and entanglement with intense light beams, we  use a linearized
approach in a rotating frame. The field mode $\hat a$ is described by $\hat
a=\alpha\hat{\openone} +\delta\hat a$. $\alpha$ describes the steady state
amplitude of the bright carrier with the frequency $\omega$ while the quantum character arising from sideband pairs is combined in the fluctuating part $\delta \hat a$. For Gaussian
states $\delta \hat a$ can be written as a superposition of the amplitude
quadrature $\delta \hat {\cal X}$ and the phase quadrature $\delta \hat {\cal Y}$: $\delta
\hat a=\frac{1}{2}(\delta \hat {\cal X}+i\delta \hat {\cal Y})$. Throughout this paper, we refer to the amplitude quadrature as the direction along the classical amplitude excitation while the phase quadrature is the noise component perpendicular to the amplitude, i.~e.\ we use a coordinate system denoted by $\delta \hat X$ and $\delta \hat Y$ that is rotated by the angle $\varphi$ (corresponding to the classical phase of the mode). We are furthermore interested in the fluctuations around the mean values ${\cal X}/2$ and ${\cal Y}/2$, hence the new coordinate system $\delta \hat X$ and $\delta \hat Y$ is displaced by ${\cal X}/2$ and ${\cal Y}/2$. For a visualization of the notation, see the inset at the bottom of fig. \ref{fig-ent}.

To characterize the quantum state of a light mode, the spectral variances of the
amplitude and the phase quadrature components $\delta \hat X$ and $\delta \hat
Y$ at a certain sideband frequency $\Omega$ are determined by spectral analysis of photocurrents. Measurements of the amplitude quadrature is easily done in direct photo detection. The amplitude quadrature is then obtained from a photon number measurement under a linear approximation assumption: $\hat n_a(t)=\hat a^\dagger \hat a=|\alpha|^2\hat{\openone}+|\alpha| \delta \hat 
X_a(t)$, where $\delta \hat X_a$ denotes the fluctuations of the amplitude quadrature 
along the classical excitation. The detection of the conjugate
variable, the phase quadrature $\delta \hat Y$, is more involved, as this
requires the introduction of a relative phase shift of the bright carrier mode
$\alpha$ with respect to the sideband modes $\delta \hat a$. A detailed analysis
of a measurement technique to obtain information about the phase quadrature
component of a light beam as well as correlation measurements of the phase
quadrature via a joint measurement on a pair of beams is the subject of section
\ref{sec-techniques}. 

Quadrature entanglement is achieved by linear interference of two intense
amplitude squeezed modes $\hat a_\mathrm{sqz}$ and $\hat
b_\mathrm{sqz}$ on a beam splitter\cite{RAL98,FUR98,LEU99}. A phase space
representation of this process is shown in figure \ref{fig-ent}. We assume that
both squeezed input beams have the same classical amplitudes
$\alpha_\mathrm{sqz}=\beta_\mathrm{sqz}=\alpha$ as well as the same variances
for the squeezed and the antisqueezed quadrature respectively: $V(\delta \hat
X_\mathrm{a,sqz})=V(\delta \hat X_\mathrm{b,sqz})\equiv V(\delta \hat
X_\mathrm{sqz})$ and $V(\delta \hat Y_\mathrm{a,sqz})=V(\delta \hat
Y_\mathrm{b,sqz}) \equiv V(\delta \hat Y_\mathrm{sqz})$. Depending on the
relative interference phase $\theta$, entangled output states $\hat
a_\mathrm{ent}$ and $\hat b_\mathrm{ent}$ are generated, with the classical
amplitudes
\begin{eqnarray}
\alpha_\mathrm{ent}&=&\alpha\sqrt{1+\cos\theta}\label{eqn-ampla}\\
\beta_\mathrm{ent}&=&\alpha\sqrt{1-\cos\theta}.\label{eqn-amplb} \end{eqnarray}
The quadrature components of the entangled output modes $\delta \hat
X_\mathrm{a,ent}$, $\delta \hat Y_\mathrm{a,ent}$, $\delta \hat
X_\mathrm{b,ent}$ and $\delta \hat Y_\mathrm{b,ent}$ are given in terms of the
input squeezed beams and the interference phase $\theta$
\begin{eqnarray}
\delta \hat X_\mathrm{a,ent}=\frac{1}{2}\frac{1}{\sqrt{1+\cos\theta}}
                [(1+\cos\theta)\delta \hat X_\mathrm{a,sqz}\nonumber \\
               +(1+\cos\theta)\delta \hat X_\mathrm{b,sqz}
                +\sin\theta\delta \hat Y_\mathrm{a,sqz}
                -\sin\theta\delta \hat Y_\mathrm{b,sqz}],\label{eqn-entxa}\\
\delta \hat Y_\mathrm{a,ent}=\frac{1}{2}\frac{1}{\sqrt{1+\cos\theta}}
                [(1+\cos\theta)\delta \hat Y_\mathrm{a,sqz}\nonumber \\
                +(1+\cos\theta)\delta \hat Y_\mathrm{b,sqz}
                -\sin\theta\delta \hat X_\mathrm{a,sqz}
                +\sin\theta\delta \hat X_\mathrm{b,sqz}],\label{eqn-entya}\\
\delta \hat X_\mathrm{b,ent}=\frac{1}{2}\frac{1}{\sqrt{1-\cos\theta}}
                [(1-\cos\theta)\delta \hat X_\mathrm{a,sqz}\nonumber \\
                +(1-\cos\theta)\delta \hat X_\mathrm{b,sqz}
                -\sin\theta\delta \hat Y_\mathrm{a,sqz}
                +\sin\theta\delta \hat Y_\mathrm{b,sqz}],\label{eqn-entxb}\\
\delta \hat Y_\mathrm{b,ent}=\frac{1}{2}\frac{1}{\sqrt{1-\cos\theta}}
                [(1-\cos\theta)\delta \hat Y_\mathrm{a,sqz}\nonumber \\
                +(1-\cos\theta)\delta \hat Y_\mathrm{b,sqz}
                +\sin\theta\delta \hat X_\mathrm{a,sqz}
                -\sin\theta\delta \hat X_\mathrm{b,sqz}].\label{eqn-entyb}
\end{eqnarray}
Pairs of quadratures are strongly correlated, in particular, for $\theta=\pi/2$
where $(\delta \hat X_\mathrm{a,ent}+\delta \hat X_\mathrm{b,ent})\rightarrow0$
and $(\delta \hat Y_\mathrm{a,ent}-\delta \hat Y_\mathrm{b,ent})\rightarrow0$
for strongly squeezed input fields. For this particular choice of the
interference phase the variances of the joint variables are
\begin{eqnarray}
V(\delta \hat
X_\mathrm{a,ent}+\delta \hat X_\mathrm{b,ent})&=&2V(\delta \hat
X_\mathrm{sqz}),\\ V(\delta \hat Y_\mathrm{a,ent}-\delta \hat
Y_\mathrm{b,ent})&=&2V(\delta \hat X_\mathrm{sqz}),
\end{eqnarray}
i.\ e.\ they only depend on the squeezing level. However, if the relative
interference phase $\theta$ between the two amplitude squeezed states is
different from $\pi/2$, the generated entanglement will not have the largest
possible value \cite{KIM02a}. In such a case the uncertainty areas in phase
space become asymmetric as illustrated in figure \ref{fig-ent}, and as a result
the correlations between quadratures of the two beams reduces.

To characterize entanglement, a well established approach is to consider the
criterion derived by Duan et al.~\cite{DUA00} and Simon \cite{SIM00}:
Entanglement or equivalently non--separability is present in the system if \cite{KOR02a}
\begin{equation}
V_\mathrm{sq}^\pm(\delta\hat X)+V_\mathrm{sq}^\mp(\delta\hat Y)<2.\label{duansimon}
\end{equation}
The correlations are expressed by the squeezing
variances
\begin{eqnarray}
V_\mathrm{sq}^\pm(\delta \hat X)&=&
\frac{V(\delta \hat X_1\pm g \delta \hat X_2)}
{V(\delta \hat X_\mathrm{1,coh} + g \delta \hat
X_\mathrm{2,coh})},\label{vquetschx}\\
V_\mathrm{sq}^\mp(\delta \hat Y)&=&
\frac{V(\delta \hat Y_1\mp g\delta \hat Y_2)}
{V(\delta \hat Y_\mathrm{1,coh} + g \delta \hat
Y_\mathrm{2,coh})}\label{vquetschy}.
\end{eqnarray}
The variances are normalized to the shot noise level of coherent states,
indicated by the subscripts "coh". Here, g is a variable gain that could be
optimized to minimize the correlation signal. For the symmetric case described
above, $g=1$. Though in general only sufficient, the criterion of inequality
(\ref{duansimon}) provides an experimentally easy measure to check whether or
not a given state is entangled. Note that by application of local unitary
squeezing operations, the sum criterion of equation (\ref{duansimon}) can be
transformed into a product criterion \cite{TAN99,GIO03,HYL05} 
\begin{equation}
V_\mathrm{sq}^\pm(\delta\hat X) \times V_\mathrm{sq}^\mp(\delta\hat Y)<1.
\end{equation}
The criterion in product form is more general than the criterion given by eqn.~(\ref{duansimon}), as a larger class of entangled states can be witnessed (in particular for biased forms of entanglement). Although the product form is in general more appropriate to verify entanglement experimentally \cite{BOW03b}, in this paper we use the sum form as it will facilitate the comparison of the different entanglement measurement schemes to be presented. As will become clear in the following, one of the measurement strategies delivers directly the sum criterion with no access to the product. 

The main purpose of this paper is to describe various methods by which a proper
linear combination of non commuting quadrature operators can be constructed to
yield the squeezing variances defined in eqns. (\ref{vquetschx}) and
(\ref{vquetschy}). Having these variances at our disposal non--separability can
subsequently be determined using the criterion (\ref{duansimon}).

\section{Experimental generation of entanglement using fiber
optics}\label{sec-experiment}

\begin{figure}
\includegraphics[scale=.45]{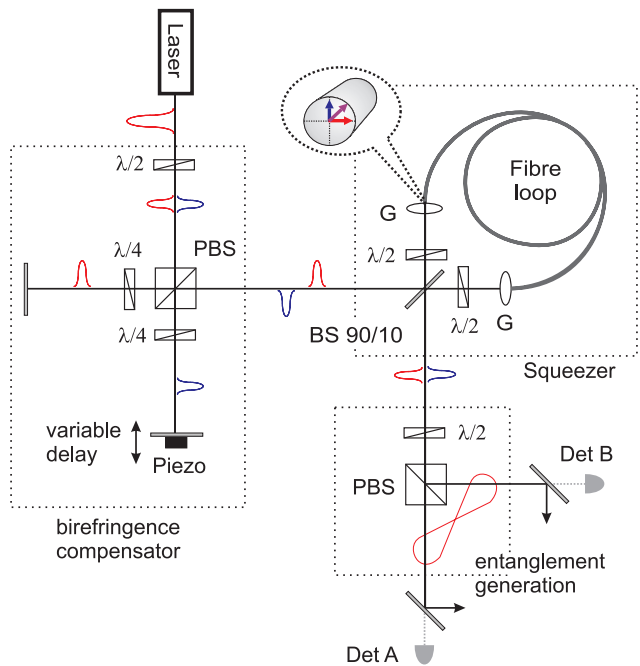}
\caption{\label{aufbau} Setup for entanglement generation. Two squeezed beams
of orthogonal polarization are generated in an asymmetric fiber Sagnac
interferometer and are brought to interference at a polarizing beam splitter
(PBS) together with a $\lambda/2$--waveplate. To control the relative phase of
the two pulses and to compensate their relative delay due to the birefringence
of the fiber, a Michelson interferometer is placed in front of the Sagnac
loop. G: Gradient index lens, Det A and B: DC--detectors for phase lock, BS:
beam splitter.} \end{figure}

In this section, we present a setup to generate intense, pulsed, amplitude
squeezed light using an asymmetric fiber Sagnac interferometer (see
Fig.~\ref{aufbau}). Intense light pulses are split into a strong and a weak
pulse at an asymmetric beam splitter. These pulses are
coupled from different directions into an optical fiber. After
counter propagating through the fiber, the pulses interfere again at the beam
splitter. Squeezing is produced by exploiting the Kerr nonlinearity: the
strong light pulses acquire an intensity dependent phase shift. Thus, the
initially circular shaped uncertainty area in phase space is transformed into an
ellipse. By interference with the weak pulse, the uncertainty ellipse is
reoriented in phase space, resulting in directly detectable amplitude squeezing
\cite{SCHM98,KRY98}.

In the experiment we use pulses with a duration of about $130$ fs at a center
wavelength of $1530$ nm and at a repetition rate of $82$ MHz. The pulses were
produced by an optical parametric oscillator (Spectra physics,OPAL) pumped by a
Ti:Sapphire Laser (Spectra Physics,Tsunami). These pulses are injected into the
asymmetric fiber Sagnac interferometer (Fig.~\ref{aufbau}). This is operated
simultaneously at two orthogonal polarizations to obtain two amplitude squeezed
beams with equal optical power for the generation of entanglement. The Sagnac
loop consists of 8~m polarization maintaining fiber (3M, FS--PM--7811) and an
asymmetric 90/10 beam splitter. In a previous experiment \cite{SIL01}, the
generation of quadrature entanglement by the interference between
the two amplitude squeezed pulses on a 50/50 beam splitter was reported. The two
orthogonally polarized squeezed modes were divided at a polarizing beam splitter
into two spatial modes. Before interference, the polarizations of both modes
were matched and the relative delay of the two pulses due to the birefringence
of the fiber was compensated.

In our experiment, the squeezed pulses
emerging the Sagnac loop in orthogonal polarization modes are already
overlapping, not only spatially but also temporally. This is achieved by
inserting a birefringence compensator in front of the Sagnac loop. This device
is basically a Michelson interferometer, that introduces different delays for
the two polarizations. Similar techniques were used by Fiorentino et al.
\cite{FIO01} to generate amplitude squeezing using a fiber in a Mach--Zehnder
like configuration and by Heersink et al.~\cite{HEE03} where polarization
squeezing was demonstrated. In order to mix the two squeezed
modes and hence to generate quadrature entanglement, a polarizing beam splitter
together with a $\lambda/2$--plate is inserted after the fiber. This allows for
the possibility to carefully align the splitting ratio of the entangling beam
splitter. To sucessfully generate entanglement, not only the delay due to the
birefringence must be compensated, but also the relative optical phase of the
two pulses must be actively locked using a feedback control system to ensure
maximum entanglement with the given resources. As noted above, this corresponds
to $\theta=\pi/2$ and hence balanced intensities of the output beams. To create
an error signal for the feed back control of the piezo mirror in the
Michelson--interferometer, the difference of the optical power of the individual
entangled beams must be recorded. This signal is accessed by detecting the
fraction of light that is leaking through high reflecting mirrors ($R>99.5$\%)
using a pair of high--gain detectors (Det A and Det B, in figure \ref{aufbau}).
The advantage of using such a system is its high stability, the almost perfect
spatial overlap of the two squeezed beams and minimized losses allowing
for an interference contrast of 99.5\%.

One specific feature of Kerr--squeezed states from optical fibers is the high
degree of excess phase noise that is acquired during propagation through the
fiber \cite{SHE90} due to guided acoustic wave Brillouin scattering (GAWBS)
\cite{SHE85a} and self phase modulation. The excess noise
in the phase quadrature was more than 20dB above the shot noise limit as
will be seen from the measurements presented below. For the generation of
entanglement, this is not a problem in principle as the phase noise leads to
classically correlated noise in the entangled output beams which cancels in the
correlation measurements. However, the phase noise does play a role when the
beam splitter for the entanglement generation is not well balanced, that is the
splitting ratio is different from 50/50. In that case, phase correlations are
degraded as the cancellation of the correlated noise is no longer perfect. In
our experiment, the excess phase noise turned out to be a limiting factor for
the entanglement generation. However, the negative effects on the correlation in
the phase quadrature was minimized by adjusting the splitting ratio of the
beam splitter properly.

\section{Experimental techniques, measurement methods and
results}\label{sec-techniques}

In the following section, we present several experimental techniques which are
capable of generating the linear combination of quadratures needed to witness
entanglement. The schemes are all based on interferometric set--ups which do not
require an external local oscillator and hence are applicable for intense,
pulsed light. Each technique described below is accompanied by experimental
results.

In all experiments, we evaluate the spectral variances of the detected light
field at a certain side band frequency $\Omega$. The detected signal is actually
the result of a beating between the bright carrier component of the light field
with a pair of quantum side bands separated symmetrically from the carrier by
$\Omega$. The spectral variance or the noise power is obtained experimentally
from the photo current of the detectors with a spectrum analyser (HP
8590). As detectors we used InGaAs photodiodes (Epitaxx ETX500).
All measurements were performed at a center frequency of $20.5$MHz or $17.5$MHz,
the resolution bandwidth was 300kHz and for averaging, a video bandwidth of 
30Hz was chosen.

\subsection{Phase measurements on intense light beams}
The most intuitive way of constructing the linear combination of quadratures
that can witness entanglement is to simultaneously measure the amplitude
quadrature (phase quadrature) of the two beams and subsequently add (subtract)
the outcomes. The amplitude quadrature of intense, pulsed light can be accessed
rather easily in direct detection with a pair of balanced detectors. However,
detection of the phase quadrature is more involved. This quadrature is normally
accessed using a homodyne detector\cite{YUE83} where the signal beam interferes
with a much stronger local oscillator. For intense signal beams, this gives rise
to technical difficulties as the high intensities may saturate the detectors.
Alternatively, phase measurements of bright beams can be achieved by
reflecting the light off a detuned cavity. A frequency dependent phase shift is
introduced due to multiple beam interference, and as a result the bright carrier
is rotated with respect to the sidebands \cite{SIE86,GAL91}. This technique was
used in some early quantum optical experiments using fibers \cite{SHE86, BAC88}.
For ultra short light pulses however, the requirements for the dispersion
properties of the resonator are quite demanding.

\begin{figure}
\includegraphics[scale=.39]{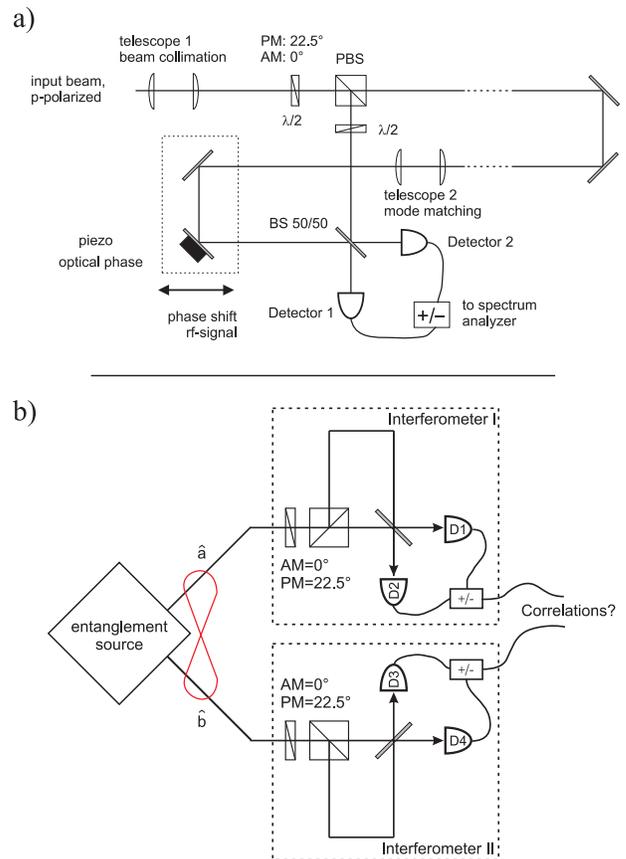}
\caption{\label{bild-interferometer} a) Experimental setup of the
phase-measuring interferometer. The orientation of the first $\lambda/2$-plate
determines the type of measurement: amplitude measurement (AM) or phase
measurement (PM). PBS, polarizing beam splitter; 50 50, beam splitter. b)
Schematic of the setup to verify entanglement using two phase measuring
interferometers.}
\end{figure}

In this section, we present another approach to measure the fluctuations of the
phase quadrature at a certain sideband frequency without the need of an external
local oscillator or a cavity. The scheme is thus suitable for the intense,
pulsed beams used in our experiment and correlations in the phase quadrature as
well as in the amplitude quadrature could be verified directly by performing
separate measurements on the entangled beam pair. Towards this goal we employed
a Mach--Zehnder interferometer with an unbalanced arm length difference
\cite{GLO04} (see Fig.~\ref{bild-interferometer}a).
In this interferometer, a phase shift between the carrier and the
sidebands (which are separated from the carrier with the frequency $\Omega$) is
introduced due to free space propagation and two beam interference. This phase
shift is achieved by splitting the input beam $\hat
a=\alpha\hat{\openone}+\delta \hat a$ into two parts and subsequently
delaying one beam with respect to the other by a length $\Delta L$. This delay
is chosen such that a phase shift of $\pi$ is acquired between the sideband
modes under investigation with respect to the carrier, corresponding to $\Delta
L=c/2f$ where $f=\Omega/2\pi$ is the measurement frequency. The beams interfere
on a second beam splitter with the relative optical phase set to $\pi/2$,
resulting in two equally bright output beams. Direct photodetection is performed
in the two output ports of the interferometer, and by taking the difference of
the photo currents, its spectral variance at the frequency $\Omega$ can be
measured. The signal is proportional to the spectral variance of the phase
quadrature of the input mode $\hat a$ at the side band frequency $\Omega$
\cite{GLO04}: \begin{equation} V(\delta \hat n_c-\delta \hat n_d)=\alpha^2
V(\delta \hat Y_a). \end{equation} Using a pulsed laser system, as in our case,
phase measurements can be performed only at certain frequencies, as possible
delays where interference occurs are governed by the repetition frequency
$f_\mathrm{rep}$ of the laser: $\Delta L= cn/f_\mathrm {rep}$ (n is an integer
number). Possible measurement frequencies are then given by
$f_m=f_\mathrm{rep}/2n$. With a repetition rate of 82 MHz, the arm length
difference thus must be  a multiple of 3.66 m, corresponding to the distance
between two successive pulses. For a measurement at a frequency of 20.5 MHz an
arm length difference of 7.32 m is required. A more detailed analysis of the
phase measuring interferometer can be found in Ref.~\cite{GLO04}. The set--up
allows for easy switching between measurements of the phase quadrature and the
amplitude quadrature. This is simply done by rotating the polarization of the
input state. Since the first beam splitter of the interferometer is a polarizing
beam splitter, one can switch between the following two operation modes: (i) the
intensity is distributed equally in both arms of the interferometer, and by
interfering the beams at the second beam splitter phase measurements are
performed (ii) all light propagates through one arm of the interferometer, the
second beam splitter together with the detector pair comprises a balanced
detector and amplitude measurements are performed.

\begin{figure}
\includegraphics[scale=.7]{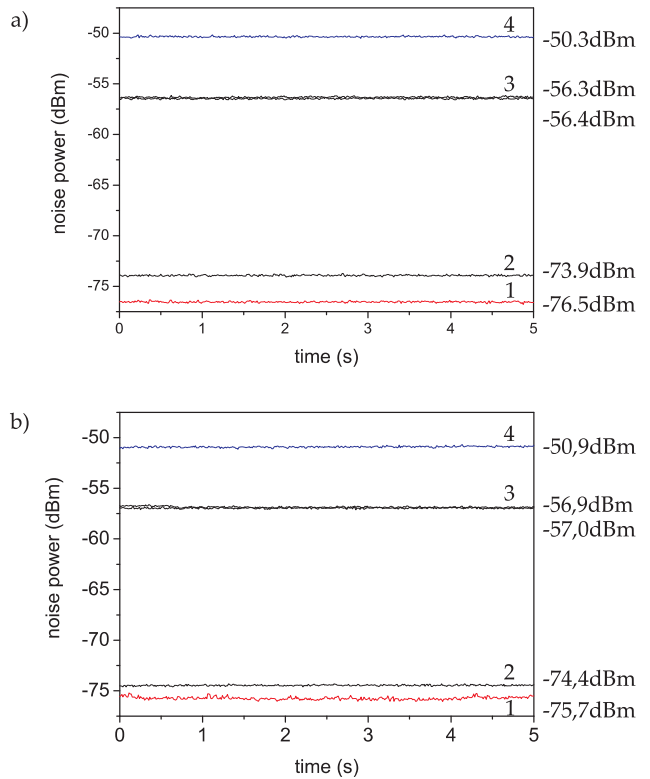}
\caption{Correlations of the (a) amplitude quadrature and (b) phase
quadrature of the entangled beam pair measured with the unbalanced
Mach--Zehnder interferometer. In each graph the noise level (traces 1) of the
correlation signal is shown together with the corresponding shot-noise level
(traces 2), the noise level of the individual beams (traces 3), and the signal
with the anticorrelations (traces 4). The traces were corrected by subtracting
the electronic noise, which was at $-84.4$ dBm. The measurement frequency was
$20.5$MHz. Similar results were presented in Ref. \cite{GLO04}, however, the
results shown here were obtained in a setup with improved
performance.}\label{ergebnisseunbalancedif} \end{figure}

To characterize entanglement, we have set up two such phase measuring
interferometers in the output ports of the entanglement setup (see
Fig.~\ref{bild-interferometer}b). First, we have investigated the input 
squeezing that is used to generate entanglement. Using the interferometric setup
in its amplitude measuring settings, we found squeezing levels of $2.5\pm0.1$dB
($2.7\pm0.1$dB) for the p (s)--polarized beam. Then, entanglement was generated
by interference of the squeezed beams. We have measured the amplitude
quadratures as well as the phase quadratures of the two entangled modes $\hat
a_\mathrm{ent}$ and $\hat b_\mathrm{ent}$ independently at different locations.
The experimental results are summarized in figure \ref{ergebnisseunbalancedif}.
In (a) traces of amplitude noise measurements of the modes $\hat a_\mathrm{ent}$
and $\hat b_\mathrm{ent}$ are plotted. Traces 3 indicate the noise levels of the
individual beams, i.~e. $V(\delta \hat X_a)$ and $V(\delta \hat X_b)$. Clearly,
both modes have the same noise level and exhibit a high degree of excess noise,
corresponding to the noisy phase quadrature component of the input squeezed
states. As discussed above, the high noise level is the result of the
antisqueezing required by the Heisenberg uncertainty relation, but also due to
additional thermal phase noise introduced by the fiber \cite{SHE85a}. The noise
level of the individual beams is more than 20dB above the quantum noise level.
However, strong anti--correlations were observed as the variance of the sum
signal of the amplitude quadratures $V(\delta \hat X_a+\delta \hat X_b)$ (trace
1) drops even below the respective shot noise level for the combined modes
(trace 2) by $2.6$dB. The squeezing variance to describe the amplitude
correlation signal therefore is given by $V_\mathrm{sq}^+(\delta \hat
X)=0.55\pm0.02$ indicating strong non--classical correlations. On the other
hand, the correlation signal, that is the difference signal of the amplitude
quadratures $V(\delta \hat X_a-\delta \hat X_b)$ is $6$dB above the noise level
of the individual beams (see trace 4). Of these $6$dB of noise, $3$dB are due to
the quantum correlations between the two modes whereas the other $3$dB comes
from the doubling in optical power. The errors indicated in this last and the
following measurements were estimated from the standard deviation of 400 values
obtained for the noise power with a spectrum analyzer at a resolution bandwidth
of $300$kHz and at a video bandwidth of $30$Hz.

In part (b) of figure \ref{ergebnisseunbalancedif}, the result of the
correlations in the phase quadrature are plotted. Again, the individual beams
exhibit a high degree of noise (traces 3), the noise level being of the same
order as that of the amplitude quadrature. Strong non--classical correlations
are observed for the phase quadrature, the variance of the difference signal
$V(\delta \hat Y_a-\delta \hat Y_b)$ (trace 1) being $1.3$dB below the quantum
noise level (trace 2). The corresponding squeezing variance is given by
$V_\mathrm{sq}^-(\delta \hat Y)=0.74\pm0.02$. The anti--correlation signal
$V(\delta \hat Y_a+\delta \hat Y_b)$ is $6$dB above the noise level of the
individual beams (trace 4). From the measured values, we can conclude that the
examined state is non--separable, as can be seen when the measured values for
the squeezing variances are plugged into eqn.~(\ref{duansimon}):
\begin{equation}
V_\mathrm{sq}^+(\delta \hat X)+V_\mathrm{sq}^-(\delta \hat
Y)=0.55+0.74=1.29\pm0.03<2.
\end{equation}
The asymmetry between the correlation measurements of the amplitude and the
phase quadrature has basically two reasons: (i) Due to the finite interference
contrast, as a result of imperfect mode matching \cite{GEA89} in the phase
measuring setup, these measurements are less efficient than amplitude
measurements, where no additional interference is required. Finite interference
contrast characterized by the visibility ${\cal V}<1$ has the effect of
introducing additional losses. The loss is related to the visibility via ${\cal
V}^2=\eta$, where $\sqrt{1-\eta}$ is the corresponding signal attenuation
of the injected mode. (ii) As discussed in the previous section,
due to slight asymmetries in the beam splitter for the entanglement generation,
the phase correlations are degraded in the presence of a high degree of excess
noise.

The major drawbacks of such an interferometer are its relative high losses due
to the long propagation distances through several optical elements and finite
mode matching quality: 10\% losses are acquired during the propagation through
the interferometer (back--reflection at lenses, losses at mirrors etc.), a
visibility of ${\cal V}=0.95$ is observed, corresponding to 10\% loss,
and the finite quantum efficiency of the photodiodes lead approximately to
another 10\% loss. The efficiency for phase quadrature measurements is therefore
roughly 73\%. In addition, when phase correlations on a pair of entangled beams
are to be detected, four detectors are involved, each of which detects
very noisy signals. Due to non--optimum balancing of the detection electronics,
a slightly reduced correlation signal is obtained. Nevertheless, this
apparatus is a very useful tool to directly measure the phase quadrature on
intense light beams. The measurements above show that sub shot noise
measurements of the phase quadrature are possible using this interferometer,
thus making it a versatile apparatus for experiments in quantum optics.
As an example, the phase measuring setup was applied in an experimental
demonstration of continuous variable quantum erasing \cite{AND04}.

\subsection{Interferometric determination of quadrature correlations}

\begin{figure}
\includegraphics[scale=.4]{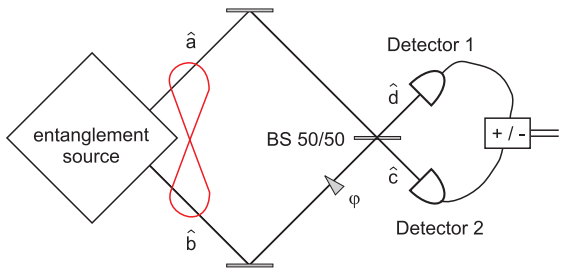} \caption{Schematic
setup for correlation measurements by interference of the entangled beam pair
at a 50/50 beam splitter and direct detection} \label{interferometrie}
\end{figure}

An alternative approach to construct the quadrature combinations that verify
the correlations between potentially entangled beams is to interfere the
entangled beam pair $\hat a_\mathrm{ent}$ and $\hat b_\mathrm{ent}$ at another
50:50 beam splitter as depicted in figure \ref{interferometrie}. Measurements of
the quantum variables are not performed on the entangled beam pair individually,
rather the joint quantum state of the beam pair is determined \cite{TAN99}.
One advantage of this interferometric scheme is that information about the
correlations in the phase quadrature can be obtained in direct detection without
additional local oscillators or other phase measuring devices, making the
experimental implementation easy. Furthermore, the cancellation of the
correlated noise is done optically thus the requirements on the detectors are
less stringent than in the previous case. Another advantage is that,
entanglement can in principle be tested in a single measurement run contrary to
the scheme in section \ref{sec-techniques}.A where two successive measurements
on the amplitude and phase quadrature must be performed.

To describe this interferometric technique, we consider the interference of the
entangled beams with relative optical phase $\varphi$ at a 50/50 beam
splitter. The output modes after the interference are given by
\begin{eqnarray}
\hat c &=& \frac{1}{\sqrt{2}}[\hat a_\mathrm{ent}
- \mathrm{e}^{i\varphi} \hat b_\mathrm{ent}]\\ \hat d &=&
\frac{1}{\sqrt{2}}[\hat a_\mathrm{ent} + \mathrm{e}^{i\varphi} \hat
b_\mathrm{ent}].
\end{eqnarray}
In both output ports of the interferometer, direct detection of the photon
number is performed. The photon numbers are given by $\hat n_c =\hat c^\dagger
\hat c$ and $\hat n_d =\hat d^\dagger \hat d$. The sum and the difference of the
fluctuating parts of the photo currents immediately deliver the correlation
signal for the amplitude and the phase quadrature of the entangled beams. For
the moment, we are interested in an interference phase of $\varphi=\pi/2$,
corresponding to equal intensity in the two output ports of the interferometer.
The correlation signals are then given by a linear combination of the quadrature
fluctuations of the entangled beam pair
\begin{eqnarray}
\delta \hat n_c+\delta \hat n_d&=&\alpha_\mathrm{ent} \delta \hat
X_\mathrm{a,ent} +\beta_\mathrm{ent} \delta \hat X_\mathrm{b,ent}\label{ifampl}\\
\delta \hat n_c-\delta \hat n_d&=&\beta_\mathrm{ent} \delta
\hat Y_\mathrm{a,ent} - \alpha_\mathrm{ent} \delta \hat Y_\mathrm{b,ent}. \label{ifphase}
\end{eqnarray}
Depending on the classical amplitudes of the entangled beam pair, the detected
correlation signal is scaled by the factors $\alpha_\mathrm{ent}$ and
$\beta_\mathrm{ent}$.

As discussed in section \ref{sec-production}, the entangled beam pair is
generated by the interference of squeezed beams of equal intensity $\alpha$ and
relative optical phase $\theta$. The amplitudes of the entangled field modes
$\hat a_\mathrm{ent}$ and $\hat b_\mathrm{ent}$ are given by
$\alpha\sqrt{1+\cos\theta}$ and $\alpha\sqrt{1-\cos\theta}$. Recall that maximal
entanglement is generated, when the interference phase $\theta$ is such that the
modes $\hat a$ and $\hat b$ have equal intensity ($\theta=\pi/2$). In that case,
the non--separability properties of the two mode state can be checked directly
in terms of the criterion by Duan and Simon \cite{DUA00,SIM00} given in
eqn.~(\ref{duansimon}). The respective values for the correlations in the
amplitude and the phase quadrature, that is the squeezing variances
$V_\mathrm{sq}^+(\hat X)$ and $V_\mathrm{sq}^-(\hat Y)$ could be found directly
using equations (\ref{ifampl}) and (\ref{ifphase}) as $\alpha=\beta$. However,
when the two potentially entangled beams have different classical amplitudes, as
in the case when the interference phase to generate entanglement is non optimum
$\theta\neq\pi/2$, non--symmetric linear combinations of the quadrature
components are measured according to equations (\ref{ifampl}) and
(\ref{ifphase}). Having such a linear combination at hand, a generalized version
of the non--separability criterion is needed as stated in \cite{BRA04,GIO03}.

We define the operators $\delta \hat u$ and $\delta \hat v$, which are linear combinations of the quadrature operators $\delta \hat X_a$ and $\delta \hat X_b$ and $\delta \hat Y_a$ and $\delta \hat Y_b$
\begin{eqnarray}
\delta \hat u&=&h_a\delta \hat X_a+ h_b\delta \hat X_b\label{eqn-lincomb1}\\
\delta \hat v&=&g_a\delta \hat Y_a+ g_b\delta \hat Y_b\label{eqn-lincomb2}.
\end{eqnarray}
Note that these new operators describe a two mode state, and the standard canonical commutation relations for single mode bosonic systems are no longer valid. However, these operators can be used to derive measureable quantities (bounderies) on non--separability as an extension to the criterion described by eqn.~(\ref{duansimon}). A generalized criterion for non--separability then reads according to references \cite{BRA04,GIO03}
\begin{equation}
\langle(\Delta \hat u)^2\rangle+\langle(\Delta \hat v)^2\rangle\leq
2(|h_a g_a|+|h_b g_b|).\label{eqn-nonsepgen}
\end{equation}
For the case described above, we easily find that
$g_a=h_b=\alpha\sqrt{1+\cos\theta}$ and $h_a=-g_b=\alpha\sqrt{1-\cos\theta}$ by
comparing (\ref{ifampl}) and (\ref{ifphase}) with (\ref{eqn-lincomb1}) and
(\ref{eqn-lincomb1}). Using these relations and normalizing to the quantum noise
limit, the criterion for non--separability adapted to our experimental
configuration can be written as
\begin{equation}
\langle(\Delta \hat u)^2\rangle_\mathrm{norm}+\langle(\Delta \hat
v)^2\rangle_\mathrm{norm}\leq2\sqrt{1-\cos^2\theta}.\label{eqn-nonsepmod}
\end{equation}
From this equation it is evident that for finite values of initial
squeezing and for interference phases $\theta\neq\pi/2$ it becomes harder to
show that the given beam pair is entangled as the right hand side of the
inequality is getting smaller. Using that criterion, only for infinite input
squeezing, entanglement is revealed for any phase $\theta\neq0$.
As long as $\theta\neq0$, entanglement is always generated with our scheme,
although it will not be witnessed by that criterion.
\begin{figure} \includegraphics[scale=.7]{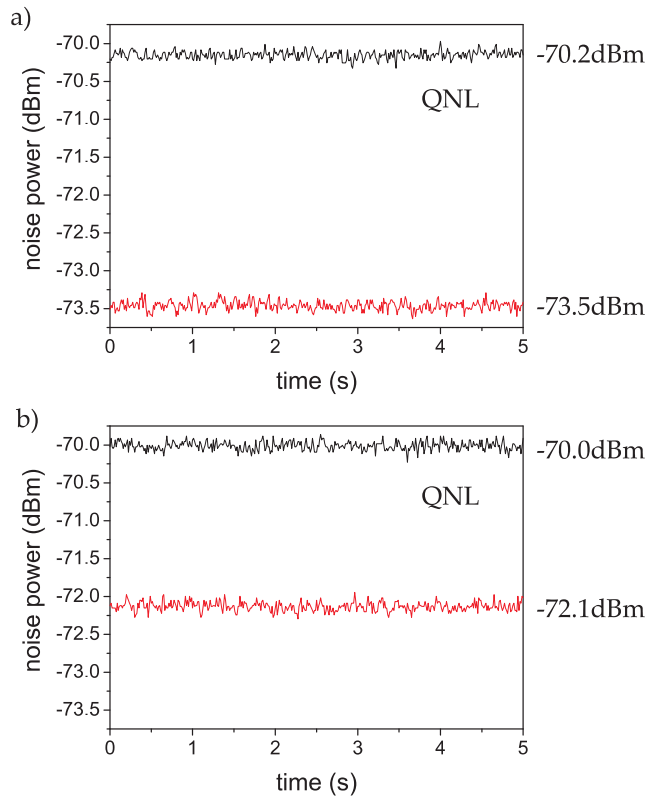}
\caption{Correlations of the (a) amplitude quadrature and (b) phase
quadrature of the entangled beam pair measured via interference and direct
detection. In the graph the noise level of the correlation signal is shown
together with the corresponding shot-noise level. The traces were corrected by
subtracting the electronic noise, which was at $-87.8$ dBm. The measurement
frequency was $17.5$MHz.} \label{bild-ergebnissekorr} \end{figure}

Interestingly, the parameters $g$ and $h$ can always be optimized such that
one can observe always as much correlations below the quantum noise limit as
there is initial squeezing. For the entangled states given in eqns.
(\ref{eqn-entxa})--(\ref{eqn-entyb}), these optimized gain factors to recover
the full correlation signal coincide with the amplitudes $\alpha_\mathrm{ent}$
and $\beta_\mathrm{ent}$ as given in eqns. (\ref{eqn-ampla}) and
(\ref{eqn-amplb}). The left hand side of inequality (\ref{eqn-nonsepmod}) can
thus be optimized to yield the same value independent of $\theta$ corresponding
to the initial squeezing level. From comparison with eqns. (\ref{ifampl}) and 
(\ref{ifphase}) we see that by using the interferometric measurement apparatus,
optimized correlation signals are measured automatically, provided the second
interference phase is chosen such that the two output modes $\hat c$ and $\hat
d$ have the same intensity. However, we point out that these correlations do not
imply maximum possible entanglement, as the boundary on the right hand side of
(\ref{eqn-nonsepmod}) has to be adjusted and the degree of violation of the
unequality changes with $\theta$.

Let us for the moment consider the case $\theta=\pi/2$, so maximum possible
entanglement is generated. Using the strategy described above, the correlations
in the amplitude and the phase quadrature are detected experimentally. The two
entangled beams interfere at a 50/50 beam splitter, the noise variance of the
sum-- and the difference signal of the photo currents are recorded. The
measurement technique used here is closely related to the one used for dense
coding by Li et al. \cite{Li02}. The results of such measurements are shown in
figure \ref{bild-ergebnissekorr}. In (a) the noise variance of the sum signal,
that is the correlation signal $V(\delta \hat X_a+\delta \hat X_b)$ of the
amplitude quadrature is plotted together with the shot noise level. The
correlation signal is $3.3$dB below the quantum noise limit, corresponding to a
squeezing variance of $V_\mathrm{sq}^+(\delta \hat X)=0.47\pm0.02$. The
correlation signal for the phase quadrature follows from the difference signal
of the photo current and is plotted in (b). The correlations $V(\delta \hat
Y_a-\delta \hat Y_b)$ of the phase quadrature are $2.1$dB below the quantum
noise limit, the squeezing variance is $V_\mathrm{sq}^-(\delta \hat
Y)=0.62\pm0.02$. From the sum of the squeezing variances \begin{equation}
V_\mathrm{sq}^+(\delta \hat X)+V_\mathrm{sq}^-(\delta \hat
Y)=0.47+0.62=1.09\pm0.03<2 \end{equation}
we can conclude the non--separability of the two mode state. Again, the
asymmetry between the sum and the difference channel, i.~e. the correlations in
the amplitude-- and the phase quadrature is due to the slightly unbalanced beam
splitting ratio in the entanglement generation scheme and in the interferometric
verification scheme together with the high degree of excess noise of the initial
squeezed states. The input squeezing used for the entanglement generation in
this and in the following experiment was also characterized with the given set
up. We measured $3.7\pm0.2$dB ($3.8\pm0.2$dB) for the p (s)--polarized input
fields. When compared to the measurement scheme where two independent phase 
measuring devices are used(Section \ref{sec-techniques}.A), this setup is more
efficient and, as expected, the detected correlations and squeezing levels are
larger.

\subsection{Direct experimental test of non--separability}

The third measurement scheme is a variation of the interferometric scheme
shown in figure \ref{interferometrie}. In contrast to
the previous measurement strategy, detection is performed in one output port
only. For a particular choice of the relative phase $\varphi$, a signal is
obtained that can be directly used to check non--separability as stated in eqn.
(\ref{duansimon}). Direct detection of the noise variances for the photon
numbers of the output modes $\hat c$ or $\hat d$ deliver the desired
correlation signal\cite{KOR02a}:
\begin{eqnarray}
        V(\delta \hat n_{c,d})
        &=&\frac{1}{2}\alpha^2(1\pm\cos\varphi)^2 V_\mathrm{sq}^+(\delta \hat X)
        \nonumber \\
        &+&\frac{1}{2}\alpha^2\sin^2\varphi
        V_\mathrm{sq}^-(\delta \hat Y).\label{eqn-duanexp}
\end{eqnarray}
The corresponding quantum noise limits are
\begin{equation}
        V(\delta \hat n_{c,d,coh})=\alpha^2(1\pm\cos\varphi).
\end{equation}
Normalizing (\ref{eqn-duanexp}) to the quantum noise limit, one obtains
\begin{eqnarray}
        &&V(\delta \hat X_{c,d})_\mathrm{norm}\nonumber \\
        &=&\frac{1}{2}\left[(1\pm\cos\varphi)V_\mathrm{sq}^+(\delta \hat X)+
        \frac{\sin^2\varphi}{(1\pm\cos\varphi)}V_\mathrm{sq}^-(\delta \hat
        Y)\right].\label{eqn-messduan}
\end{eqnarray}
For the interference phase $\varphi=\pi/2$ the detected signal in one output
mode reduces to
\begin{equation}
V(\delta \hat X_{c,d})_\mathrm{norm}=\frac{1}{2}\left[
V_\mathrm{sq}^+(\delta \hat X)+
V_\mathrm{sq}^-(\delta \hat Y) \right]
\end{equation}
which is exactly the measure one needs to prove non--separability according to
the Duan--Simon criterion. The two--mode state is therefore entangled, if noise
reduction in the output ports $\hat c$ or $\hat d$ are observed.

\begin{figure}
\includegraphics[scale=.7]{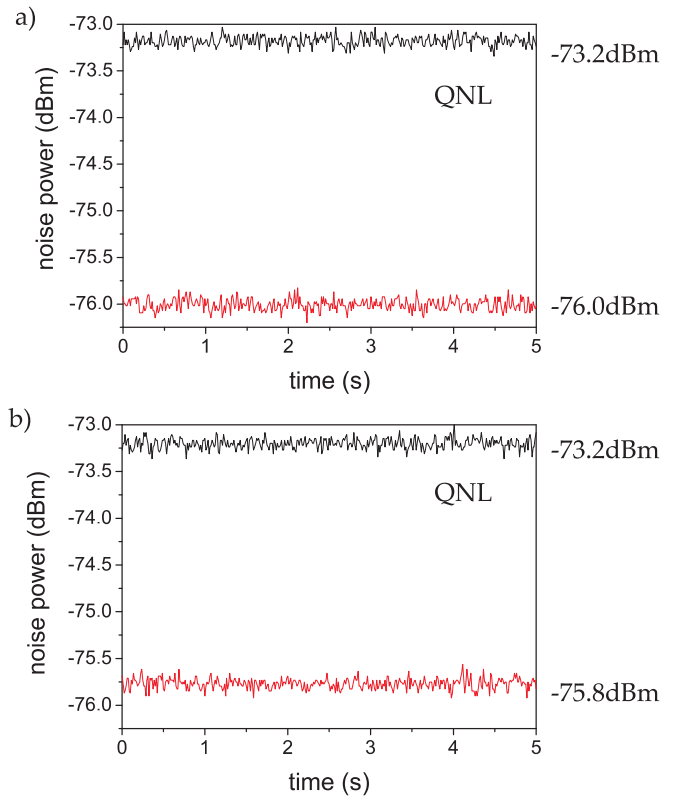}
\caption{Direct detection of non--separability. Both entangled beams were
brought to interference, the noise power is measured in the individual output
ports. Amplitude squeezing of $2.8$dB and $2.6$dB below the
shot noise level is observed. the measurement frequency
was $17.5$MHz, the electronic noise level was at
$-87.8$dBm.}\label{bild-ergebnisseduan} \end{figure}

The results of such
measurements on modes $\hat c$ and $\hat d$ are plotted in figure
\ref{bild-ergebnisseduan}. The recorded noise level of the modes is plotted
together with the shot noise level. Both output modes are squeezed, and we
measure $V(\delta \hat X_{c})_\mathrm{norm}=0.52\pm0.02$ (see figure
\ref{bild-ergebnisseduan} (a)) and $V(\delta \hat
X_{d})_\mathrm{norm}=0.55\pm0.02$ (see \ref{bild-ergebnisseduan} (b)). From
these noise levels we obtain the sum of the squeezing variances
\begin{eqnarray}
V_\mathrm{sq}^+(\delta \hat X)+ V_\mathrm{sq}^-(\delta \hat
Y)&=&1.05\pm0.04<2,\\
V_\mathrm{sq}^+(\delta \hat X)+ V_\mathrm{sq}^-(\delta
\hat Y)&=&1.10\pm0.04<2, \end{eqnarray}
for the measurements in the output ports $\hat c$ and $\hat d$, respectively.
The values are in good agreement with the measurement results of the previous
section, where the correlations of the amplitude and the phase quadrature have
been determined separately by using the sum-- and the difference channel of both
output ports $\hat c$ and $\hat d$. The results shown in
Fig.~\ref{bild-ergebnisseduan} were obtained by an improved version of the
measurement reported by Silberhorn et al. \cite{SIL01}. In contrast to earlier
measurements, the two phases $\theta$ and $\varphi$ are now stabilized and, most
importantly, the phase $\theta$ was adjusted to $\pi/2$ so that the
non--separability criterion holds in its standard form (see eqns.
(\ref{duansimon}) and (\ref{eqn-nonsepmod})).

To conclude from such measurements that the beam pair under investigation is
indeed entangled when noise reduction in the output modes is observed, requires
that both beams have equal intensity. Then the Duan--Simon criterion in the form
(\ref{duansimon}) can be applied directly to check for non--separability. Note
that in this case squeezing is observed in the two output ports of the
interference for any value of $\varphi$ as can be seen from equation
(\ref{eqn-messduan}). For all other values of $\theta$, i.\ e.\
$\theta\neq\pi/2$, more general criteria have to be applied to check for
entanglement, like those given in equation (\ref{eqn-nonsepgen}). In this
situation however, the expression (\ref{eqn-messduan}) to describe the noise
power in one output port of the interferometer as a function of the interference
phase $\varphi$ looks more complicated.

We summarize and compare the results of the different methods
to characterize entanglement of intense light beams in
Table \ref{table-comparison}.
\begin{table}[h!]
\begin{ruledtabular}
\begin{tabular}{clc}
   &Method& $V_\mathrm{sq}^+(\delta\hat X)+V_\mathrm{sq}^-(\delta \hat Y)$\\
\hline
A. & Phase measuring &\\
   &interferometer   &  \raisebox{1.5ex}[-1.5ex]{$1.29\pm0.03$} \\
B. & Measure correlations &\\
   &interferometrically & \raisebox{1.5ex}[-1.5ex]{$1.09\pm0.03$}\\
C. & Direct test of &$1.05\pm0.04$ (output $\hat c$)\\
   &non--separability & $1.10\pm0.04$ (output $\hat d$)
\end{tabular}
\end{ruledtabular}
\caption{Comparison of the different methods to verify entanglement of intense,
pulsed light. For each method, the experimental results for the inseparability
criterion $V_\mathrm{sq}^+(\delta\hat X)+V_\mathrm{sq}^-(\delta \hat Y)<2$ as
stated in eqn.~(\ref{duansimon}) are noted. }\label{table-comparison}
\end{table}
All methods deliver roughly the same results, and we note in particular that
their estimated errors are of the same order. However, method A cannot be
compared directly with the other two methods (B and C) since the measurement
frequency was $20.5$MHz in the former case while it was $17.5$MHz in the latter
case. The slightly poorer results of method A can be explained easily by its
higher intrinsic losses that reduce the degree of quantum correlations.
Depending on the type of applications and whether or not it is possible to bring
the entangled beams to interference, one can choose the most appropriate way to
access information about the quantum correlations of a beam pair.

\section{Conclusions}\label{sec-conclusions}

In this paper, we have presented a comprehensive study of a fiber optical
entanglement source with bright, pulsed light. The source is rather compact,
robust and easy to handle. Not only entanglement generation was shown, but
also different measurement tools were provided to detect entanglement. Hence
these methods can be readily implemented in protocols for quantum
communication and quantum information processing. These schemes all rely on
interference effects, thus they are suited to handle intense beams as they do
not require external local oscillators. Stable operation and the generation of
correlations well beyond the quantum noise limit was demonstrated. For a certain
class of applications, such as entanglement swapping or quantum dense coding
with continuous variables, the high degree of excess noise introduced in the
fibers, limits the efficiency and performance of these protocols. However, the
problem of excess noise might be overcome by using microstructured fibers for
the squeezing generation. New experiments indicate that less thermal noise is
generated when these fibers are used instead of standard fibers
\cite{KORN04}. Less thermal noise would also improve the measurement accuracy,
as the required dynamic range to be handled in the detection electronics could
be considerably reduced. Together with general improvements in the detection
electronics this gives potential for improved entanglement generation and
detection using the fiber setup. The whole setup could be simplified further by
changing to a fiber integrated Sagnac loop for the squeezed state generation, as
was demonstrated in reference \cite{MEI04}.

\begin{acknowledgments}
We want to acknowledge helpful discussions with P.~van Loock. This work was
supported by the Schwer\-punkt\-pro\-gramm 1078 of the Deutsche
Forschungsgemeinschaft (QIV), the Kompetenznetzwerk
Quanteninformationsverarbeitung of the state of Bavaria (A8) and the EU project
COVAQIAL (project no. FP6-511004).
\end{acknowledgments}

%\bibliography{literaturliste}% Produces the bibliography via BibTeX.

\begin{thebibliography}{42}
\expandafter\ifx\csname natexlab\endcsname\relax\def\natexlab#1{#1}\fi
\expandafter\ifx\csname bibnamefont\endcsname\relax
  \def\bibnamefont#1{#1}\fi
\expandafter\ifx\csname bibfnamefont\endcsname\relax
  \def\bibfnamefont#1{#1}\fi
\expandafter\ifx\csname citenamefont\endcsname\relax
  \def\citenamefont#1{#1}\fi
\expandafter\ifx\csname url\endcsname\relax
  \def\url#1{\texttt{#1}}\fi
\expandafter\ifx\csname urlprefix\endcsname\relax\def\urlprefix{URL }\fi
\providecommand{\bibinfo}[2]{#2}
\providecommand{\eprint}[2][]{\url{#2}}

\bibitem[{\citenamefont{Einstein et~al.}(1935)\citenamefont{Einstein, Podolsky,
  and Rosen}}]{EIN35}
\bibinfo{author}{\bibfnamefont{A.}~\bibnamefont{Einstein}},
  \bibinfo{author}{\bibfnamefont{B.}~\bibnamefont{Podolsky}}, \bibnamefont{and}
  \bibinfo{author}{\bibfnamefont{N.}~\bibnamefont{Rosen}},
  \bibinfo{journal}{Physical Review} \textbf{\bibinfo{volume}{47}},
  \bibinfo{pages}{777} (\bibinfo{year}{1935}).

\bibitem[{\citenamefont{Nielsen and Chuang}(2000)}]{NIE00}
\bibinfo{author}{\bibfnamefont{M.~A.} \bibnamefont{Nielsen}} \bibnamefont{and}
  \bibinfo{author}{\bibfnamefont{I.~L.} \bibnamefont{Chuang}},
  \emph{\bibinfo{title}{Quantum {C}omputation and {Q}uantum {I}nformation}}
  (\bibinfo{publisher}{Cambridge {U}niversity {P}ress},
  \bibinfo{address}{Cambridge}, \bibinfo{year}{2000}).

\bibitem[{\citenamefont{Bennett et~al.}(1993)\citenamefont{Bennett, Brassard,
  Crépeau, Jozsa, Peres, and Wootters}}]{BEN93}
\bibinfo{author}{\bibfnamefont{C.~H.}~\bibnamefont{Bennett}},
  \bibinfo{author}{\bibfnamefont{G.}~\bibnamefont{Brassard}},
  \bibinfo{author}{\bibfnamefont{C.}~\bibnamefont{Crepeau}},
  \bibinfo{author}{\bibfnamefont{R.}~\bibnamefont{Jozsa}},
  \bibinfo{author}{\bibfnamefont{A.}~\bibnamefont{Peres}}, \bibnamefont{and}
  \bibinfo{author}{\bibfnamefont{W.~K.}~\bibnamefont{Wootters}},
  \bibinfo{journal}{Physical Review Letters} \textbf{\bibinfo{volume}{70}},
  \bibinfo{pages}{1895} (\bibinfo{year}{1993}).

\bibitem[{\citenamefont{Bennett and Wiesner}(1992)}]{BEN92}
\bibinfo{author}{\bibfnamefont{C.~H.} \bibnamefont{Bennett}} \bibnamefont{and}
  \bibinfo{author}{\bibfnamefont{S.~J.} \bibnamefont{Wiesner}},
  \bibinfo{journal}{Physical Review Letters} \textbf{\bibinfo{volume}{69}},
  \bibinfo{pages}{2881} (\bibinfo{year}{1992}).

\bibitem[{\citenamefont{Braunstein and Pati}(2003)}]{BRA03}
\bibinfo{editor}{\bibfnamefont{S.~L.} \bibnamefont{Braunstein}}
  \bibnamefont{and} \bibinfo{editor}{\bibfnamefont{A.~K.} \bibnamefont{Pati}},
  eds., \emph{\bibinfo{title}{Quantum Information with Continuous Variables}}
  (\bibinfo{publisher}{Kluwer Akademic Press}, \bibinfo{address}{Dordrecht},
  \bibinfo{year}{2003}).

\bibitem[{\citenamefont{Ou et~al.}(1992)\citenamefont{Ou, Pereira, Kimble, and
  Peng}}]{OU92}
\bibinfo{author}{\bibfnamefont{Z.~Y.}~\bibnamefont{Ou}},
  \bibinfo{author}{\bibfnamefont{S.~F.}~\bibnamefont{Pereira}},
  \bibinfo{author}{\bibfnamefont{H.~J.}~\bibnamefont{Kimble}}, \bibnamefont{and}
  \bibinfo{author}{\bibfnamefont{K.~C.}~\bibnamefont{Peng}},
  \bibinfo{journal}{Physical Review Letters} \textbf{\bibinfo{volume}{68}},
  \bibinfo{pages}{3663} (\bibinfo{year}{1992}).

\bibitem[{\citenamefont{Zhang et~al.}(2000)\citenamefont{Zhang, Wang, Li, Jing,
  Xie, and Peng}}]{ZHA00b}
\bibinfo{author}{\bibfnamefont{Y.}~\bibnamefont{Zhang}},
  \bibinfo{author}{\bibfnamefont{H.}~\bibnamefont{Wang}},
  \bibinfo{author}{\bibfnamefont{X.} \bibnamefont{Li}},
  \bibinfo{author}{\bibfnamefont{J.} \bibnamefont{Jing}},
  \bibinfo{author}{\bibfnamefont{C.} \bibnamefont{Xie}}, \bibnamefont{and}
  \bibinfo{author}{\bibfnamefont{K.} \bibnamefont{Peng}},
  \bibinfo{journal}{Physical Review A} \textbf{\bibinfo{volume}{62}},
  \bibinfo{pages}{023813} (\bibinfo{year}{2000}).

\bibitem[{\citenamefont{Laurat et~al.}(2004)\citenamefont{Laurat, Coudreau,
  Keller, Treps, and Fabre}}]{LAU04}
\bibinfo{author}{\bibfnamefont{J.}~\bibnamefont{Laurat}},
  \bibinfo{author}{\bibfnamefont{T.}~\bibnamefont{Coudreau}},
  \bibinfo{author}{\bibfnamefont{G.}~\bibnamefont{Keller}},
  \bibinfo{author}{\bibfnamefont{N.}~\bibnamefont{Treps}}, \bibnamefont{and}
  \bibinfo{author}{\bibfnamefont{C.}~\bibnamefont{Fabre}},
  \bibinfo{journal}{Phys. Rev. A} \textbf{\bibinfo{volume}{70}},
  \bibinfo{pages}{042315} (\bibinfo{year}{2004}).

\bibitem[{\citenamefont{Schori et~al.}(2002)\citenamefont{Schori, S{\o}rensen,
  and Polzik}}]{SCHO02}
\bibinfo{author}{\bibfnamefont{C.}~\bibnamefont{Schori}},
  \bibinfo{author}{\bibfnamefont{J.~L.} \bibnamefont{S{\o}rensen}},
  \bibnamefont{and} \bibinfo{author}{\bibfnamefont{E.~S.}
  \bibnamefont{Polzik}}, \bibinfo{journal}{Physical Review A}
  \textbf{\bibinfo{volume}{66}}, \bibinfo{pages}{033802}
  (\bibinfo{year}{2002}).

\bibitem{VIL05}
\bibinfo{author}{\bibfnamefont{A.~S.}~\bibnamefont{Villar}},
  \bibinfo{author}{\bibfnamefont{L.~S.} \bibnamefont{Cruz}},
  \bibinfo{author}{\bibfnamefont{K.~N.} \bibnamefont{Cassemiro}},
  \bibinfo{author}{\bibfnamefont{M.} \bibnamefont{Martinelli}},
  \bibnamefont{and} \bibinfo{author}{\bibfnamefont{P.}
  \bibnamefont{Nussenzveig}},
  \bibinfo{pages}{arXiv:quant--ph/0506139}
  (\bibinfo{year}{2002}).

\bibitem[{\citenamefont{Wenger et~al.}(2004)\citenamefont{Wenger, Ourjoumtsev,
  Tualle-Brouri, and Grangier}}]{WEN05}
\bibinfo{author}{\bibfnamefont{J.}~\bibnamefont{Wenger}},
  \bibinfo{author}{\bibfnamefont{A.}~\bibnamefont{Ourjoumtsev}},
  \bibinfo{author}{\bibfnamefont{R.}~\bibnamefont{Tualle-Brouri}},
  \bibnamefont{and} \bibinfo{author}{\bibfnamefont{F.}~\bibnamefont{Grangier}},
  \bibinfo{journal}{Eur. Phys. J. D} \textbf{\bibinfo{volume}{32}},
  \bibinfo{pages}{391} (\bibinfo{year}{2004}).

\bibitem[{\citenamefont{Furusawa et~al.}(1998)\citenamefont{Furusawa,
  S{\o}rensen, Braunstein, Fuchs, Kimble, and Polzik}}]{FUR98}
\bibinfo{author}{\bibfnamefont{A.}~\bibnamefont{Furusawa}},
  \bibinfo{author}{\bibfnamefont{J.}~\bibnamefont{S{\o}rensen}},
  \bibinfo{author}{\bibfnamefont{S.}~\bibnamefont{Braunstein}},
  \bibinfo{author}{\bibfnamefont{C.}~\bibnamefont{Fuchs}},
  \bibinfo{author}{\bibfnamefont{H.}~\bibnamefont{Kimble}}, \bibnamefont{and}
  \bibinfo{author}{\bibfnamefont{E.}~\bibnamefont{Polzik}},
  \bibinfo{journal}{Science} \textbf{\bibinfo{volume}{282}},
  \bibinfo{pages}{706} (\bibinfo{year}{1998}).

\bibitem[{\citenamefont{Bowen et~al.}(2003)\citenamefont{Bowen, Schnabel, Lam,
  and Ralph}}]{BOW03b}
\bibinfo{author}{\bibfnamefont{W.~P.} \bibnamefont{Bowen}},
  \bibinfo{author}{\bibfnamefont{R.}~\bibnamefont{Schnabel}},
  \bibinfo{author}{\bibfnamefont{P.~K.} \bibnamefont{Lam}}, \bibnamefont{and}
  \bibinfo{author}{\bibfnamefont{T.~C.} \bibnamefont{Ralph}},
  \bibinfo{journal}{Physical Review Letters} \textbf{\bibinfo{volume}{90}},
  \bibinfo{pages}{043601} (\bibinfo{year}{2003}).

\bibitem[{\citenamefont{Aoki et~al.}(2003)\citenamefont{Aoki, Takei, Yonezawa,
  Wakui, Hiraoka, Furusawa, and van Loock}}]{AOK03}
\bibinfo{author}{\bibfnamefont{T.}~\bibnamefont{Aoki}},
  \bibinfo{author}{\bibfnamefont{N.}~\bibnamefont{Takei}},
  \bibinfo{author}{\bibfnamefont{H.}~\bibnamefont{Yonezawa}},
  \bibinfo{author}{\bibfnamefont{K.}~\bibnamefont{Wakui}},
  \bibinfo{author}{\bibfnamefont{T.}~\bibnamefont{Hiraoka}},
  \bibinfo{author}{\bibfnamefont{A.}~\bibnamefont{Furusawa}}, \bibnamefont{and}
  \bibinfo{author}{\bibfnamefont{P.}~\bibnamefont{van Loock}},
  \bibinfo{journal}{Physical Review Letters} \textbf{\bibinfo{volume}{91}},
  \bibinfo{pages}{080404} (\bibinfo{year}{2003}).

\bibitem[{\citenamefont{Josse et~al.}(2004)\citenamefont{Josse, Dantan,
  Bramati, Pinard, and Giacobino}}]{JOS04}
\bibinfo{author}{\bibfnamefont{V.}~\bibnamefont{Josse}},
  \bibinfo{author}{\bibfnamefont{A.}~\bibnamefont{Dantan}},
  \bibinfo{author}{\bibfnamefont{A.}~\bibnamefont{Bramati}},
  \bibinfo{author}{\bibfnamefont{M.}~\bibnamefont{Pinard}}, \bibnamefont{and}
  \bibinfo{author}{\bibfnamefont{E.}~\bibnamefont{Giacobino}},
  \bibinfo{journal}{Physical Review Letters} \textbf{\bibinfo{volume}{92}},
  \bibinfo{pages}{123601} (\bibinfo{year}{2004}).

\bibitem[{\citenamefont{Silberhorn et~al.}(2001)\citenamefont{Silberhorn, Lam,
  Wei{\ss}, K\"onig, Korolkova, and Leuchs}}]{SIL01}
\bibinfo{author}{\bibfnamefont{C.}~\bibnamefont{Silberhorn}},
  \bibinfo{author}{\bibfnamefont{P.~K.} \bibnamefont{Lam}},
  \bibinfo{author}{\bibfnamefont{O.}~\bibnamefont{Wei{\ss}}},
  \bibinfo{author}{\bibfnamefont{F.}~\bibnamefont{K\"onig}},
  \bibinfo{author}{\bibfnamefont{N.}~\bibnamefont{Korolkova}},
  \bibnamefont{and} \bibinfo{author}{\bibfnamefont{G.}~\bibnamefont{Leuchs}},
  \bibinfo{journal}{Physical Review Letters} \textbf{\bibinfo{volume}{86}},
  \bibinfo{pages}{4267} (\bibinfo{year}{2001}).

\bibitem[{\citenamefont{Reid and Drummond}(1988)}]{REI88}
\bibinfo{author}{\bibfnamefont{M.~D.} \bibnamefont{Reid}} \bibnamefont{and}
  \bibinfo{author}{\bibfnamefont{P.~D.} \bibnamefont{Drummond}},
  \bibinfo{journal}{Physical Review Letters} \textbf{\bibinfo{volume}{60}},
  \bibinfo{pages}{2731} (\bibinfo{year}{1988}).

\bibitem[{\citenamefont{Duan et~al.}(2000)\citenamefont{Duan, Giedke, Cirac,
  and Zoller}}]{DUA00}
\bibinfo{author}{\bibfnamefont{L.~M.}~\bibnamefont{Duan}},
  \bibinfo{author}{\bibfnamefont{G.}~\bibnamefont{Giedke}},
  \bibinfo{author}{\bibfnamefont{J.~I.}~\bibnamefont{Cirac}}, \bibnamefont{and}
  \bibinfo{author}{\bibfnamefont{P.}~\bibnamefont{Zoller}},
  \bibinfo{journal}{Physical Review Letters} \textbf{\bibinfo{volume}{84}},
  \bibinfo{pages}{2722} (\bibinfo{year}{2000}).

\bibitem[{\citenamefont{Simon}(2000)}]{SIM00}
\bibinfo{author}{\bibfnamefont{R.}~\bibnamefont{Simon}},
  \bibinfo{journal}{Physical Review Letters} \textbf{\bibinfo{volume}{84}},
  \bibinfo{pages}{2726} (\bibinfo{year}{2000}).

\bibitem[{\citenamefont{Gl\"ockl et~al.}(2004)\citenamefont{Gl\"ockl, Andersen,
  Lorenz, Silberhorn, Korolkova, and Leuchs}}]{GLO04}
\bibinfo{author}{\bibfnamefont{O.}~\bibnamefont{Gl\"ockl}},
  \bibinfo{author}{\bibfnamefont{U.~L.} \bibnamefont{Andersen}},
  \bibinfo{author}{\bibfnamefont{S.}~\bibnamefont{Lorenz}},
  \bibinfo{author}{\bibfnamefont{C.}~\bibnamefont{Silberhorn}},
  \bibinfo{author}{\bibfnamefont{N.}~\bibnamefont{Korolkova}},
  \bibnamefont{and} \bibinfo{author}{\bibfnamefont{G.}~\bibnamefont{Leuchs}},
  \bibinfo{journal}{Optics Letters} \textbf{\bibinfo{volume}{29}},
  \bibinfo{pages}{1936} (\bibinfo{year}{2004}).

\bibitem[{\citenamefont{Ralph and Lam}(1998)}]{RAL98}
\bibinfo{author}{\bibfnamefont{T.~C.}~\bibnamefont{Ralph}} \bibnamefont{and}
  \bibinfo{author}{\bibfnamefont{P.~K.}~\bibnamefont{Lam}},
  \bibinfo{journal}{Physical Review Letters} \textbf{\bibinfo{volume}{81}},
  \bibinfo{pages}{5668} (\bibinfo{year}{1998}).

\bibitem[{\citenamefont{Leuchs et~al.}(1999)\citenamefont{Leuchs, Ralph,
  Silberhorn, and Korolkova}}]{LEU99}
\bibinfo{author}{\bibfnamefont{G.}~\bibnamefont{Leuchs}},
  \bibinfo{author}{\bibfnamefont{T.}~\bibnamefont{Ralph}},
  \bibinfo{author}{\bibfnamefont{C.}~\bibnamefont{Silberhorn}},
  \bibnamefont{and}
  \bibinfo{author}{\bibfnamefont{N.}~\bibnamefont{Korolkova}},
  \bibinfo{journal}{Journal of Modern Optics} \textbf{\bibinfo{volume}{46}},
  \bibinfo{pages}{1927} (\bibinfo{year}{1999}).

\bibitem[{\citenamefont{Kim et~al.}(2002)\citenamefont{Kim, Son, Buzek, and
  Knight}}]{KIM02a}
\bibinfo{author}{\bibfnamefont{M.~S.} \bibnamefont{Kim}},
  \bibinfo{author}{\bibfnamefont{W.}~\bibnamefont{Son}},
  \bibinfo{author}{\bibfnamefont{V.}~\bibnamefont{Buzek}}, \bibnamefont{and}
  \bibinfo{author}{\bibfnamefont{P.~L.} \bibnamefont{Knight}},
  \bibinfo{journal}{Phys. Rev. A} \textbf{\bibinfo{volume}{65}},
  \bibinfo{pages}{032323} (\bibinfo{year}{2002}).

\bibitem[{\citenamefont{Korolkova et~al.}(2002)\citenamefont{Korolkova,
  Silberhorn, Gl\"ockl, Lorenz, Marquardt, and Leuchs}}]{KOR02a}
\bibinfo{author}{\bibfnamefont{N.}~\bibnamefont{Korolkova}},
  \bibinfo{author}{\bibfnamefont{C.}~\bibnamefont{Silberhorn}},
  \bibinfo{author}{\bibfnamefont{O.}~\bibnamefont{Gl\"ockl}},
  \bibinfo{author}{\bibfnamefont{S.}~\bibnamefont{Lorenz}},
  \bibinfo{author}{\bibfnamefont{C.}~\bibnamefont{Marquardt}},
  \bibnamefont{and} \bibinfo{author}{\bibfnamefont{G.}~\bibnamefont{Leuchs}},
  \bibinfo{journal}{European Physical Journal D} \textbf{\bibinfo{volume}{18}},
  \bibinfo{pages}{229} (\bibinfo{year}{2002}).

\bibitem[{\citenamefont{Tan}(1999)}]{TAN99}
\bibinfo{author}{\bibfnamefont{S.~M.} \bibnamefont{Tan}},
  \bibinfo{journal}{Physical Review A} \textbf{\bibinfo{volume}{60}},
  \bibinfo{pages}{2752} (\bibinfo{year}{1999}).

\bibitem[{\citenamefont{Giovannetti et~al.}(2003)\citenamefont{Giovannetti,
  Mancini, Vitali, and Tombesi}}]{GIO03}
\bibinfo{author}{\bibfnamefont{V.}~\bibnamefont{Giovannetti}},
  \bibinfo{author}{\bibfnamefont{S.}~\bibnamefont{Mancini}},
  \bibinfo{author}{\bibfnamefont{D.}~\bibnamefont{Vitali}}, \bibnamefont{and}
  \bibinfo{author}{\bibfnamefont{P.}~\bibnamefont{Tombesi}},
  \bibinfo{journal}{Physical Review A} \textbf{\bibinfo{volume}{67}}
  \bibinfo{pages}{022320}(\bibinfo{year}{2003}).
  
\bibitem{HYL05}
\bibinfo{author}{\bibfnamefont{P.}~\bibnamefont{Hyllus}},
  \bibnamefont{and}
  \bibinfo{author}{\bibfnamefont{J.}~\bibnamefont{Eisert}},
  \bibinfo{pages}{arXiv:quant--ph/0510077} (\bibinfo{year}{2005}).  

\bibitem[{\citenamefont{Schmitt et~al.}(1998)\citenamefont{Schmitt, Ficker,
  Wolff, K\"onig, Sizmann, and Leuchs}}]{SCHM98}
\bibinfo{author}{\bibfnamefont{S.}~\bibnamefont{Schmitt}},
  \bibinfo{author}{\bibfnamefont{J.}~\bibnamefont{Ficker}},
  \bibinfo{author}{\bibfnamefont{M.}~\bibnamefont{Wolff}},
  \bibinfo{author}{\bibfnamefont{F.}~\bibnamefont{K\"onig}},
  \bibinfo{author}{\bibfnamefont{A.}~\bibnamefont{Sizmann}}, \bibnamefont{and}
  \bibinfo{author}{\bibfnamefont{G.}~\bibnamefont{Leuchs}},
  \bibinfo{journal}{Physical Review Letters} \textbf{\bibinfo{volume}{81}},
  \bibinfo{pages}{2446} (\bibinfo{year}{1998}).

\bibitem[{\citenamefont{Krylov and Bergman}(1998)}]{KRY98}
\bibinfo{author}{\bibfnamefont{D.}~\bibnamefont{Krylov}} \bibnamefont{and}
  \bibinfo{author}{\bibfnamefont{K.}~\bibnamefont{Bergman}},
  \bibinfo{journal}{Optics Letters} \textbf{\bibinfo{volume}{23}},
  \bibinfo{pages}{1390} (\bibinfo{year}{1998}).

\bibitem[{\citenamefont{Fiorentino et~al.}(2001)\citenamefont{Fiorentino,
  Sharping, Kumar, Levandovsky, and Vasilyev}}]{FIO01}
\bibinfo{author}{\bibfnamefont{M.}~\bibnamefont{Fiorentino}},
  \bibinfo{author}{\bibfnamefont{J.~E.} \bibnamefont{Sharping}},
  \bibinfo{author}{\bibfnamefont{P.}~\bibnamefont{Kumar}},
  \bibinfo{author}{\bibfnamefont{D.}~\bibnamefont{Levandovsky}},
  \bibnamefont{and} \bibinfo{author}{\bibfnamefont{M.}~\bibnamefont{Vasilyev}},
  \bibinfo{journal}{Physical Review A} \textbf{\bibinfo{volume}{64}},
  \bibinfo{pages}{031801(R)} (\bibinfo{year}{2001}).

\bibitem[{\citenamefont{Heersink et~al.}(2003)\citenamefont{Heersink, Gaber,
  Lorenz, Gl\"ockl, Korolkova, and Leuchs}}]{HEE03}
\bibinfo{author}{\bibfnamefont{J.}~\bibnamefont{Heersink}},
  \bibinfo{author}{\bibfnamefont{T.}~\bibnamefont{Gaber}},
  \bibinfo{author}{\bibfnamefont{S.}~\bibnamefont{Lorenz}},
  \bibinfo{author}{\bibfnamefont{O.}~\bibnamefont{Gl\"ockl}},
  \bibinfo{author}{\bibfnamefont{N.}~\bibnamefont{Korolkova}},
  \bibnamefont{and} \bibinfo{author}{\bibfnamefont{G.}~\bibnamefont{Leuchs}},
  \bibinfo{journal}{Phys. Rev. A} \textbf{\bibinfo{volume}{68}},
  \bibinfo{pages}{013815} (\bibinfo{year}{2003}).

\bibitem[{\citenamefont{Shelby et~al.}(1990)\citenamefont{Shelby, Drummond, and
  Carter}}]{SHE90}
\bibinfo{author}{\bibfnamefont{R.~M.} \bibnamefont{Shelby}},
  \bibinfo{author}{\bibfnamefont{P.~D.} \bibnamefont{Drummond}},
  \bibnamefont{and} \bibinfo{author}{\bibfnamefont{S.~J.}
  \bibnamefont{Carter}}, \bibinfo{journal}{Physical Review A}
  \textbf{\bibinfo{volume}{42}}, \bibinfo{pages}{2966} (\bibinfo{year}{1990}).

\bibitem[{\citenamefont{Shelby et~al.}(1985)\citenamefont{Shelby, Levenson, and
  Bayer}}]{SHE85a}
\bibinfo{author}{\bibfnamefont{R.~M.} \bibnamefont{Shelby}},
  \bibinfo{author}{\bibfnamefont{M.~D.} \bibnamefont{Levenson}},
  \bibnamefont{and} \bibinfo{author}{\bibfnamefont{P.~W.} \bibnamefont{Bayer}},
  \bibinfo{journal}{Physical Review Letters} \textbf{\bibinfo{volume}{54}},
  \bibinfo{pages}{939} (\bibinfo{year}{1985}).

\bibitem[{\citenamefont{Yuen and Chan}(1983)}]{YUE83}
\bibinfo{author}{\bibfnamefont{H.~P.} \bibnamefont{Yuen}} \bibnamefont{and}
  \bibinfo{author}{\bibfnamefont{V.}~\bibnamefont{Chan}},
  \bibinfo{journal}{Optics Letters} \textbf{\bibinfo{volume}{8}},
  \bibinfo{pages}{177} (\bibinfo{year}{1983}).

\bibitem[{\citenamefont{Siegman}(1986)}]{SIE86}
\bibinfo{author}{\bibfnamefont{A.~E.} \bibnamefont{Siegman}},
  \emph{\bibinfo{title}{Lasers}} (\bibinfo{publisher}{University Science
  Books}, \bibinfo{address}{Mill Valley}, \bibinfo{year}{1986}).

\bibitem[{\citenamefont{Galatola et~al.}(1991)\citenamefont{Galatola, Lugiato,
  G., Tombesi, and Leuchs}}]{GAL91}
\bibinfo{author}{\bibfnamefont{P.}~\bibnamefont{Galatola}},
  \bibinfo{author}{\bibfnamefont{L.~A.} \bibnamefont{Lugiato}},
  \bibinfo{author}{\bibfnamefont{P.~M.} \bibnamefont{G.}},
  \bibinfo{author}{\bibfnamefont{P.}~\bibnamefont{Tombesi}}, \bibnamefont{and}
  \bibinfo{author}{\bibfnamefont{G.}~\bibnamefont{Leuchs}},
  \bibinfo{journal}{Optics Communications} \textbf{\bibinfo{volume}{85}},
  \bibinfo{pages}{95} (\bibinfo{year}{1991}).

\bibitem[{\citenamefont{Shelby et~al.}(1986)\citenamefont{Shelby, Levenson,
  Perlmutter, DeVoe, and Walls}}]{SHE86}
\bibinfo{author}{\bibfnamefont{R.~M.} \bibnamefont{Shelby}},
  \bibinfo{author}{\bibfnamefont{M.~D.} \bibnamefont{Levenson}},
  \bibinfo{author}{\bibfnamefont{S.~H.} \bibnamefont{Perlmutter}},
  \bibinfo{author}{\bibfnamefont{R.~G.} \bibnamefont{DeVoe}}, \bibnamefont{and}
  \bibinfo{author}{\bibfnamefont{D.~F.} \bibnamefont{Walls}},
  \bibinfo{journal}{Phys. Rev. Lett.} \textbf{\bibinfo{volume}{57}},
  \bibinfo{pages}{691} (\bibinfo{year}{1986}).

\bibitem[{\citenamefont{Bachor et~al.}(1988)\citenamefont{Bachor, Levenson,
  Walls, Perlmutter, and Shelby}}]{BAC88}
\bibinfo{author}{\bibfnamefont{H.~A.} \bibnamefont{Bachor}},
  \bibinfo{author}{\bibfnamefont{M.~D.} \bibnamefont{Levenson}},
  \bibinfo{author}{\bibfnamefont{D.~F.} \bibnamefont{Walls}},
  \bibinfo{author}{\bibfnamefont{S.~H.} \bibnamefont{Perlmutter}},
  \bibnamefont{and} \bibinfo{author}{\bibfnamefont{R.~M.}
  \bibnamefont{Shelby}}, \bibinfo{journal}{Phys. Rev. A}
  \textbf{\bibinfo{volume}{38}}, \bibinfo{pages}{180} (\bibinfo{year}{1988}).

\bibitem[{\citenamefont{Gea-Banacloche and Leuchs}(1989)}]{GEA89}
\bibinfo{author}{\bibfnamefont{J.}~\bibnamefont{Gea-Banacloche}}
  \bibnamefont{and} \bibinfo{author}{\bibfnamefont{G.}~\bibnamefont{Leuchs}},
  \bibinfo{journal}{Journal of Modern Optics} \textbf{\bibinfo{volume}{36}},
  \bibinfo{pages}{1277} (\bibinfo{year}{1989}).

\bibitem[{\citenamefont{Andersen et~al.}(2004)\citenamefont{Andersen, Gl\"ockl,
  Lorenz, Leuchs, and Filip}}]{AND04}
\bibinfo{author}{\bibfnamefont{U.~L.} \bibnamefont{Andersen}},
  \bibinfo{author}{\bibfnamefont{O.}~\bibnamefont{Gl\"ockl}},
  \bibinfo{author}{\bibfnamefont{S.}~\bibnamefont{Lorenz}},
  \bibinfo{author}{\bibfnamefont{G.}~\bibnamefont{Leuchs}}, \bibnamefont{and}
  \bibinfo{author}{\bibfnamefont{R.}~\bibnamefont{Filip}},
  \bibinfo{journal}{Phys. Rev. Lett.} \textbf{\bibinfo{volume}{93}},
  \bibinfo{pages}{100403} (\bibinfo{year}{2004}).

\bibitem[{\citenamefont{Braunstein and van Loock}(2004)}]{BRA04}
\bibinfo{author}{\bibfnamefont{S.~L.} \bibnamefont{Braunstein}}
  \bibnamefont{and} \bibinfo{author}{\bibfnamefont{P.}~\bibnamefont{van
  Loock}}, \bibinfo{journal}{Rev. Mod. Phys.}
\textbf{\bibinfo{volume}{77}}, \bibinfo{pages}{513} (\bibinfo{year}{2005}).

\bibitem[{\citenamefont{Li et~al.}(2002)\citenamefont{Li, Pan, Jing, Zhang,
  Xie, and Peng}}]{Li02}
\bibinfo{author}{\bibfnamefont{X.}~\bibnamefont{Li}},
  \bibinfo{author}{\bibfnamefont{Q.}~\bibnamefont{Pan}},
  \bibinfo{author}{\bibfnamefont{J.}~\bibnamefont{Jing}},
  \bibinfo{author}{\bibfnamefont{J.}~\bibnamefont{Zhang}},
  \bibinfo{author}{\bibfnamefont{C.}~\bibnamefont{Xie}}, \bibnamefont{and}
  \bibinfo{author}{\bibfnamefont{K.}~\bibnamefont{Peng}},
  \bibinfo{journal}{Physical Review Letters} \textbf{\bibinfo{volume}{88}},
  \bibinfo{pages}{047904} (\bibinfo{year}{2002}).

\bibitem[{\citenamefont{Korn et~al.}()\citenamefont{Korn, Gl\"ockl, Lorenz,
  Marquardt, Andersen, and Leuchs}}]{KORN04}
\bibinfo{author}{\bibfnamefont{A.}~\bibnamefont{Korn}},
  \bibinfo{author}{\bibfnamefont{O.}~\bibnamefont{Gl\"ockl}},
  \bibinfo{author}{\bibfnamefont{S.}~\bibnamefont{Lorenz}},
  \bibinfo{author}{\bibfnamefont{C.}~\bibnamefont{Marquardt}},
  \bibinfo{author}{\bibfnamefont{U.~L.} \bibnamefont{Andersen}},
  \bibnamefont{and} \bibinfo{author}{\bibfnamefont{G.}~\bibnamefont{Leuchs}},
CLEO/IQEC and PhAST Technical Digest on CDROM (The Optical Society of   America,
Washington, DC, 2004), ITuI37.

\bibitem[{\citenamefont{Mei{\ss}ner et~al.}(2004)\citenamefont{Mei{\ss}ner,
  Marquardt, Heersink, Gaber, Wietfeld, Leuchs, and Andersen}}]{MEI04}
\bibinfo{author}{\bibfnamefont{M.}~\bibnamefont{Mei{\ss}ner}},
  \bibinfo{author}{\bibfnamefont{C.}~\bibnamefont{Marquardt}},
  \bibinfo{author}{\bibfnamefont{J.}~\bibnamefont{Heersink}},
  \bibinfo{author}{\bibfnamefont{T.}~\bibnamefont{Gaber}},
  \bibinfo{author}{\bibfnamefont{A.}~\bibnamefont{Wietfeld}},
  \bibinfo{author}{\bibfnamefont{G.}~\bibnamefont{Leuchs}}, \bibnamefont{and}
  \bibinfo{author}{\bibfnamefont{U.~L.} \bibnamefont{Andersen}},
  \bibinfo{journal}{J. Opt. B: Quantum Semiclass. Opt.}
  \textbf{\bibinfo{volume}{6}}, \bibinfo{pages}{S652} (\bibinfo{year}{2004}).

\end{thebibliography}

\end{document}